\documentclass[symmetry,article,accept,pdftex,moreauthors]{Definitions/mdpi} 

\firstpage{1} 
\pubvolume{15}
\issuenum{1}
\articlenumber{1948}
\pubyear{2023}
\copyrightyear{2023}
\externaleditor{Academic Editor: Bhawna Gomber}
\datereceived{15 September 2023} 
\daterevised{12 October 2023} 
\dateaccepted{17 October 2023} 
\datepublished{20 October 2023 } 
\hreflink{https://\linebreak doi.org/10.3390/sym15101948} 
\usepackage{caption}
\usepackage[labelformat=simple]{subcaption}

\DeclareCaptionLabelFormat{subcaptionlabel}{\normalfont(\textbf{#2}\normalfont)}
\newcommand{\JP}{$J/\psi$~}
\newcommand{\Ds}{$D_s$~}
\newcommand{\Dm}{$D$~}

\Title{Identifying \Dm Mesons from Radiative $W$ Decays at the Large Hadron Collider}

\TitleCitation{Identifying \Dm Mesons from Radiative $W$ Decays at the Large Hadron Collider}

\Author{Evelin Bakos 
 $^{1,2,*}$\orcidA{}, Nicolo de Groot $^{2}$\orcidB{} and Nenad Vranjes $^{1}$\orcidC{}}

\AuthorNames{Evelin Bakos, Nicolo de Groot and Nenad Vranjes}

\AuthorCitation{Bakos, E.; de Groot, N.; Vranjes, N.} 

\address{%
$^{1}$ \quad Institute of Physics, University of Belgrade, Pregrevica 118, 11080 Belgrade, Serbia; nenad.vranjes@cern.ch
\\
$^{2}$ \quad Institute for Mathematics, Astrophysics and Particle Physics (IMAPP), Radboud University and Nikhef, Heyendaalseweg 135, 
6525 AJ Nijmegen, The Netherlands; 
n.degroot@science.ru.nl}

\corres{Correspondence: evelin.bakos@cern.ch}

\abstract{In this paper, we present two machine learning algorithms to identify \Dm mesons produced in a colour singlet state from radiative $W$ boson decays at the LHC. The combined network algorithm is able to identify \Dm mesons via its hadronic decays with an efficiency of 47\% while suppressing a background of quark and gluon jets by a factor of 100. Using the developed algorithm, we perform a prospective 
	study for the measurement of $\mathcal{B}(W\to D_{s}\gamma)$.}

\keyword{standard model QCD processes; W-boson rare decays; jet tagging; machine learning} 

\begin{document}

\section{Introduction}
\label{sec:intro}

The large amount of $W$ bosons produced in $pp$ collisions at the Large Hadron Collider (LHC) enables searches for exclusive hadronic decays. These decay modes can offer precision studies of QCD factorisation \cite{Grossmann:2015lea} and are sensitive to the coupling of the $W$ boson with the photon. However, the searches for the hadronic decays are still challenging due to the large background dominated by various QCD processes. 
Of all these decay modes, $W\to D_s\gamma$ has the largest branching fraction predicted by the Standard Model with the value of $\mathcal{B} = (3.7\pm1.5)\times 10^{-8}$. No such decay has been observed so far and the best upper limit is set by the LHCb collaboration with the value $\mathcal{B}(W\to D_{s}\gamma) < 6.4\times10^{-4}$ at a 95\% confidence level \cite{LHCb:2022kta}. The limit is obtained by analysing $K^+K^-\pi^+$ final states, which make up 5.4\% of the \Ds decays. This improves on an earlier limit of $\mathcal{B}(W\to D_{s}\gamma) < 1.3\times10^{-3}$ set by the CDF collaboration \cite{Abe:1998vm}, using only $\phi(K^+K^-)\pi^+$ and $K^{*0}K^+$ final states, which comprise 3.9\% of all \Ds decays. The algorithm presented in this paper offers a new approach to identify \Dm  mesons specific to radiative $W$ 
boson decay, using inclusive tagging, and is sensitive to all decays at the possible expense of higher backgrounds. As a proof of principle, we focus on the \Ds meson because of its highest predicted branching ratio.

A recent study \cite{10.21468/SciPostPhysCore.2.2.008} demonstrated that jets originating from a radiative decay of a colour-singlet charmonium state can be distinguished from coloured jets. With machine learning algorithms, we can differentiate between jets originating from radiatively  produced \Ds mesons and background jets from gluons and quarks. The main characteristic is that they are produced without accompanying fragmentation tracks and produce isolated jets. With retraining, the algorithm offers an opportunity to identify other mesons originating from hadronic decays of colour-singlet states as well. This would improve future searches for these rare decays and could improve the measurement precision using data to be collected during the ongoing LHC Run 3.

In the following section, the simulation setup is described together with the algorithm where a deep neural network (DNN), a convolutional neural network (CNN), and a combined network are used to identify signal mesons. In Section~\ref{sec3}, the results are presented and an overview is given of the network performance and stability. In Section~\ref{sec4}, prospects for the search for $W\to D_s\gamma$ are assessed.  
\section{Materials and Methods}

\subsection{Simulated Samples}

Proton--proton collisions are simulated at 13.6 TeV to match the Run 3 data taking period of the LHC. 
The sample of \Ds particles is obtained via the hadronic decay of the $W\to D_s \gamma$. The matrix element for the process $pp\to W$ is generated at LO accuracy in QCD using \texttt{MADGRAPHv5}~\cite{Alwall:2007st,Alwall:2014hca}. The \texttt{NN23LO1} PDF set~\cite{NNPDF:2017mvq} is used in the generation. The $W\to D_s\gamma$ decay as well as parton showering and subsequent hadronisation are performed using  \texttt{Pythia8}~\cite{Sjostrand:2007gs} with the A14 ATLAS tune~\cite{ATL-PHYS-PUB-2014-021}.

The main background processes (in terms of $D_{s}$ identification) are $pp\to gg$ and $pp\to qq$ where $g$ and $q$ denotes a gluon and a quark, respectively. The background samples are modelled separately using the same setup as for the signal events. 


The detector response is simulated via the \texttt{Delphes}~\cite{deFavereau:2013fsa} package using the ATLAS detector configuration files. Jets are reconstructed as pFlow jets with the anti-$k_t$ \cite{Cacciari:2008gp} jet clustering algorithm with $\Delta R =$ 0.4, and are required to satisfy $p_T >$25~GeV and $|\eta| <$~2.1 selection criteria. Jets are considered as a \Ds meson if the angular distance to the truth \Ds particle is $\Delta R<0.2$. The entire configuration can be found in \cite{git}.

\subsection{\texorpdfstring{\Ds}{DS} Identification using Machine Learning Algorithm}

The full set of $W\to D_s\gamma$ signal sample consists of 180k events, the $qq$ background sample contains 45k, and the $gg$ background sample contains 30k events. In addition, $qq$ and $gg$, 30k, 30k and 45k $Z\to \Upsilon/(J/\psi)/\phi+\gamma$ events are considered as background to ensure that the network is able to reject other colour-singlet states. 
This makes the full background sample with 160k events comparable to the signal. Before the training all the samples were divided into training and testing sets, consisting of 70\% and 30\% of the full dataset, respectively. To create the machine learning algorithm, TensorFlow \cite{DBLP:journals/corr/AbadiBCCDDDGIIK16} and Keras \cite{chollet2015keras} libraries were used. To determine the model performance, we use the receiver operating characteristic (ROC) curve and in particular the area under the ROC curve (AuC). The network hyperparameters were optimised with grid search to make sure that the best performing models were used to obtain the results.

\subsubsection{Deep Neural Network}

Signal jets originate from the decay of an isolated \Ds (not surrounded by fragmentation tracks) and will be more
collimated than background jets. This is particularly true for gluon jets, since gluon-initiated jets have higher particle multiplicity and a softer fragmentation function, due to the large colour factor. Variables $\Delta\phi$ and $\Delta\eta$, which measure the width of the jet in the $\phi$ and $\eta$ directions, as well as $R_{em}$ and $R_{track}$ which measure the $\Delta R$ with respect to the jet axis in case of tracks and electromagnetic clusters can be used to distinguish jets originating from \Ds from the background jets. The multiplicity of charged and neutral particles ($n_{ch}$ and $n_0$) in jets  originating from \Ds is lower compared to jets from quarks and gluons. From the lower constituent multiplicity it can also be deducted that signal jets have lower invariant mass. The $m_{tr}$ measures the invariant mass of all charged tracks while $m_j$ defines the invariant mass of all constituents in the jet. Jets emerging from \Ds mesons are also less surrounded with hadronic activity caused by the fragmentation. The $p_{core}$ and $f_{core}$ measure the  ratio of sum $p_T$ in a cone  and the jet $p_T$, and the ratio of sum $E_T$ in a cone  and the jet total $E_T$, respectively. The $E_{had}/E_{em}$ defines the energy ratio in the hadronic and the electromagnetic calorimeter.

We start with the variables used in
\cite{10.21468/SciPostPhysCore.2.2.008}. These variables are further extended with the absolute values of the total charge and the jet-charge ($p_T$ weighted charge sum \cite{Field:1977fa}). The charge is expected to peak at zero for gluon jets, at one for signal jets, and have a higher average value for  quark jets. In addition, with the b-jet tagging we gain some discriminating power against b-jets. A particular class of generalised angularities ($\lambda^k_\beta$) \cite{Larkoski:2014pca} are also added to the algorithm, which are efficient in distinguishing quark jets from gluon jets.

\begin{figure}[H]
	\includegraphics[width=\textwidth]{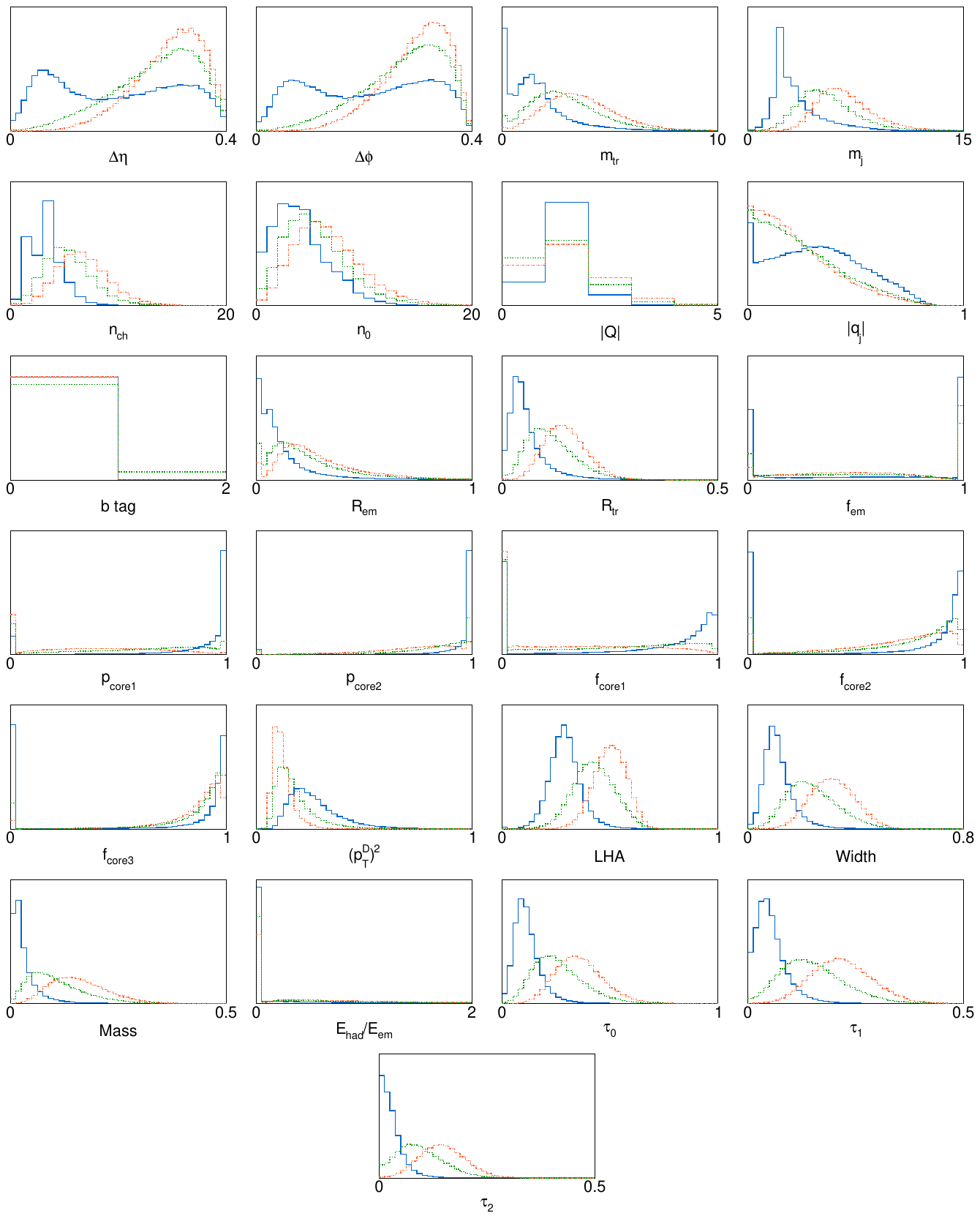}
	\caption{Distributions 
	of the variables used for \Ds identification, using DNN. The signal is presented with a solid blue line, while the $gg$ and $qq$ backgrounds are drawn with dashed red and dotted green lines, respectively.}
	\label{fig:ds_variables}
\end{figure}

Furthermore, the N-Subjettiness \cite{Thaler:2010tr} is also used,
which measures to what degree the jet is composed of N subjets. For our signal jets, the N-subjettiness was expected to be close to zero, since all the radiation is aligned with the direction of the jet, meaning N (or fewer) subjets. $gg$ background jets have $\tau_N >>$ 0, since a large fraction of their energy is distributed away from the jet direction. All the variables used for the ML algorithm are listed in Table~\ref{tab:ds_variables} and also shown in Figure \ref{fig:ds_variables}. 

\begin{table}[H]
        \caption{DNN input variables.}
	\begin{tabularx}{\textwidth}{ll}
		\toprule
		\textbf{Name} & \textbf{Description} \\
		\midrule
		$\Delta\eta$ & width of the jet in $\eta$\\
		$\Delta\phi$ & width of the jet in $\phi$ \\
		$m_{tr}$ & invariant mass of all charged tracks in the jet \\
		$m_j$ & invariant mass of all constituents of the jet \\ 
		$n_{ch}$ & charged particle multiplicity \\ 
		$n_{0}$ & neutral particle multiplicity \\ 
		$|Q|$ & absolute value of the total charge \\
		$|q_j|$ & jet charge ($p_T$ weighted charge sum, $\Sigma_i\; q_i\cdot p_{Ti}^{1/2} / \Sigma_i\; p_{Ti}^{1/2}$) \\
		$b$-tag & output of the $b$-tagging algorithm \\ 
		$R_{em}$ & average $\Delta R$ with respect to the jet axis weighted by electromagnetic energy \\
		$R_{track}$ & $p_T$ weighted average $\Delta R$ for tracks \\
		$f_{em}$ & fraction of EM energy over total neutral energy of the jet \\
		$p_{core1}$ & ratio of sum $p_T$ in a cone of $\Delta R < $0.1 and the jet $p_T$ \\
		$p_{core2}$ & ratio of sum $p_T$ in a cone of $\Delta R < $0.2 and the jet $p_T$ \\
		$f_{core1}$ & ratio of sum ET in a cone of $\Delta R <$ 0.1 and the jet total ET \\
		$f_{core2}$ & ratio of sum ET in a cone of $\Delta R <$ 0.2 and the jet total ET \\
		$f_{core3}$ & ratio of sum ET in a cone of $\Delta R <$ 0.3 and the jet total ET \\
		$(p_T^D)^2$ & $\lambda^2_{0}$ \\
		LHA & Les Houches Angularity; $\lambda^1_{0.5}$ \\
		Width & $\lambda^1_{1}$ \\
		Mass & $\lambda^1_{2}$ \\
		$E_{had}/E_{em}$ & ratio of the hadronic versus electromagnetic energy deposited in the calorimeter \\ 
		$\tau_0$, $\tau_1$, $\tau_2$ & N-Subjettiness \\
		\bottomrule
	\end{tabularx}
	\label{tab:ds_variables}
\end{table}
\unskip

\begin{figure}[H]
     \includegraphics[width=0.5\textwidth]{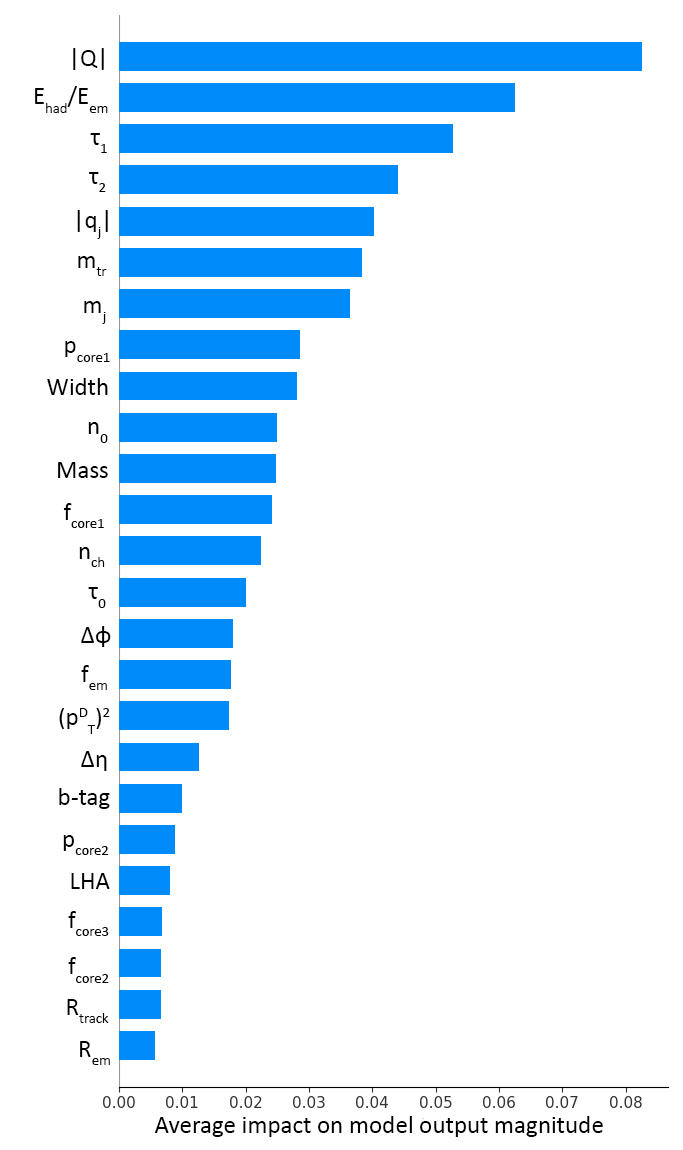}
     \caption{Feature importance plot of DNN. The blue bars represent the weight of each feature (variable) within the network.}
     \label{fig:features}
\end{figure}

Based on the optimisation results, the final model consists of one input layer and two hidden layers with 35, 20 and 12 nodes, respectively. The activation function for the input layer and both hidden layers is \texttt{tanh}. As is common with classification problems, the output layer is activated with the \texttt{sigmoid} function. The full set of hyperparameters is summarised in Table~\ref{tab:model}. A feature importance plot for the DNN network is also presented in Figure~\ref{fig:features}. It can be seen that the most important features are the charge ratio of the hadronic and electromagnetic energy deposit, and the N-Subjettiness, while the $R_{em}$ variable has very little impact on the network performance. This indicates kinematics of the generated sample does not have a major impact on the obtained results.

\unskip
\begin{table}[H]
        \caption{Hyperparameters of the different network types.} 
	\begin{adjustwidth}{-\extralength}{0cm}
        \begin{tabularx}{\fulllength}{L C C C}
		\toprule
		\textbf{Parameter} & \textbf{DNN} & \textbf{CNN} & \textbf{Combined} \\
		\midrule
		Dense layer nodes & 35---20---12---1 & -- & 33---20---14 \\ 
        Dense layer activation & tanh---tanh---tanh---sigmoid & -- & tanh---tanh---tanh \\
        \midrule
		Convolutional layer nodes & -- &  30---8---8 & 30---8---8 \\
		Window size & -- & [3 $\times$ 3], [3 $\times$ 3], [5 $\times$ 5] & [3 $\times$ 3], [3 $\times$ 3], [5 $\times$ 5]  \\
		Convolutional layer activation & -- & tanh---tanh---tanh & tanh---tanh---tanh \\
		Max pooling & -- & \multicolumn{2}{c}{After the 1st convolutional layer} \\
		Dense layers after convolution & -- & 10(relu)---1(sigmoid) & -- \\
		\midrule
		Combined layer nodes & -- & -- & 8---1\\ 
		Combined layer activation & -- & -- & relu---sigmoid\\ 
		\midrule
		Loss function & \multicolumn{3}{c}{binary cross-entropy} \\
		Optimiser & \multicolumn{3}{c}{Adam} \\
		Training epochs & \multicolumn{3}{c}{40} \\
		Batch size & \multicolumn{3}{c}{1024} \\
		\bottomrule
	\end{tabularx}
        \end{adjustwidth}
	\label{tab:model}
\end{table}
\subsubsection{Convolutional Neural Network}

Another approach for developing a \Ds identification algorithm is to use a CNN. In this case the input variables are low level variables: energy deposited in the electromagnetic and the hadronic calorimeter, and track transverse momentum, which are plotted as a 3D image. The advantage of this approach is that one can use relatively raw data instead of carefully constructed variables.

\begin{figure}[H]
     \begin{subfigure}[b]{0.45\textwidth}
         \includegraphics[width=0.45\textwidth]{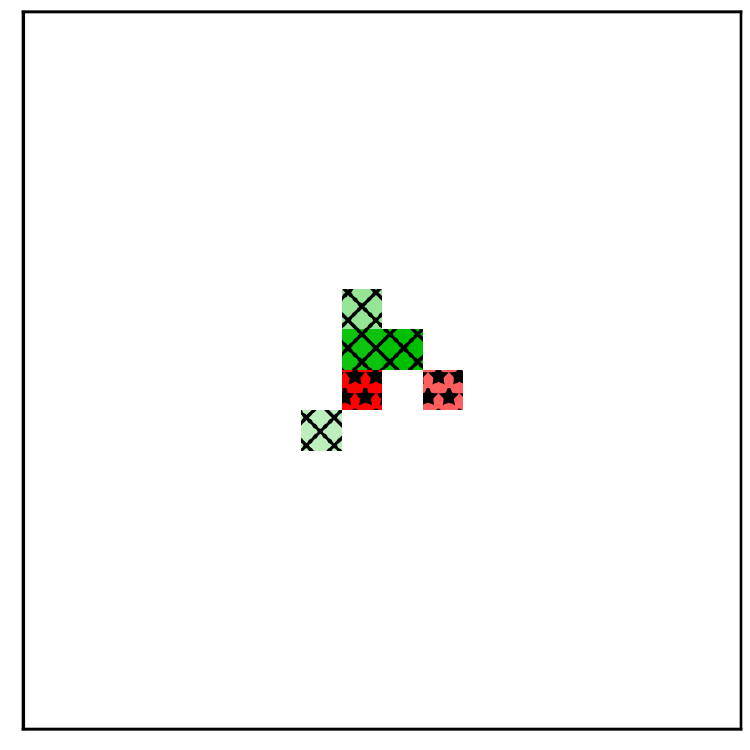}
         \includegraphics[width=0.45\textwidth]{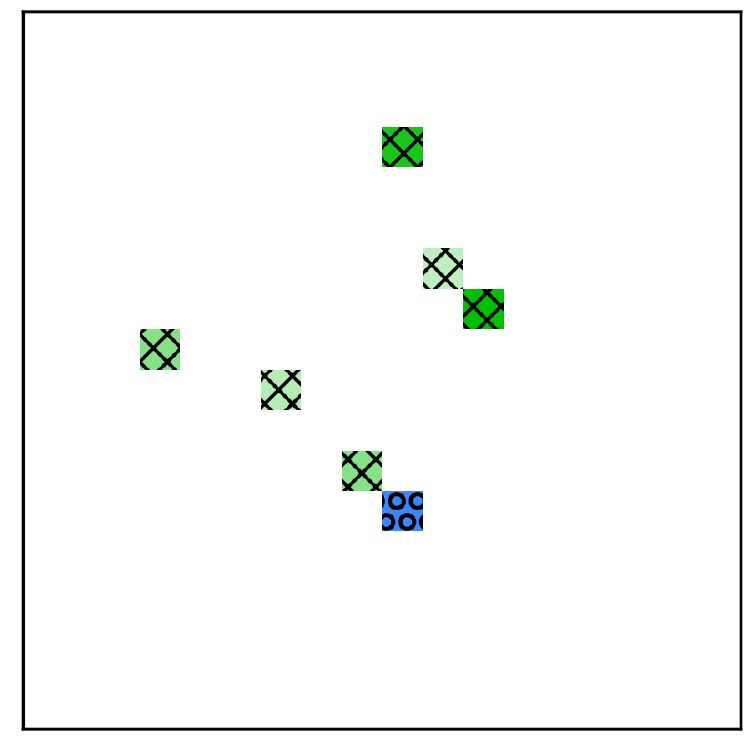}
         \caption{\centering}
         \label{fig:cnn_dsgamma_sig}
     \end{subfigure}
     \begin{subfigure}[b]{0.45\textwidth}
         \includegraphics[width=0.45\textwidth]{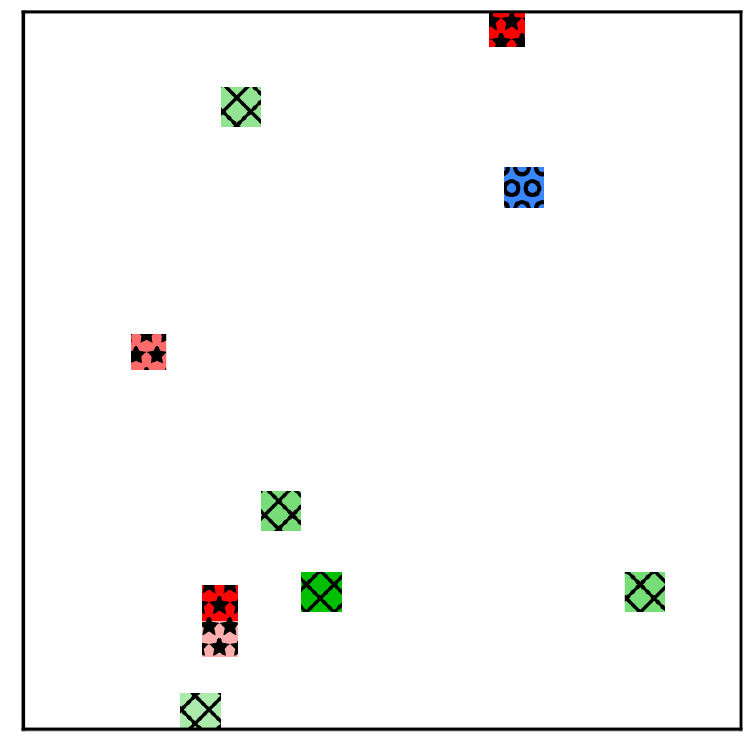}
         \includegraphics[width=0.45\textwidth]{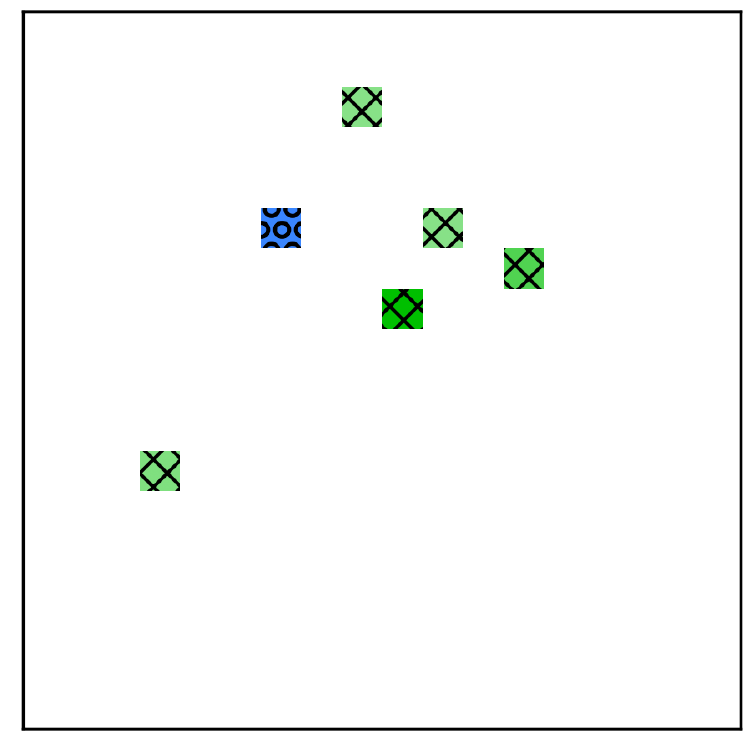}
         \caption{\centering}
         \label{fig:cnn_dsgamma_bkg}
     \end{subfigure}
     \caption{Jet image construction from low level variables, where (\textbf{a}) shows 2 different signal and (\textbf{b}) 2 different background events. The hadronic deposit is noted with blue circle pattern, the electromagnetic deposit with green square grid and the track transverse momentum with red star pattern composing an RGB input picture to the CNN algorithm.}
     \label{fig:cnn_dsgamma}
\end{figure}

In the context of this analysis, these energy deposits and the track transverse momentum are converted into a 20 $\times$ 20 jet image. Since the jet reconstruction parameter is $\Delta R =$ 0.4, and the segmentation of the ECAL is 0.02 $\times$ 0.02, the grid size of the jet image is equal to the smallest possible tower size in the $\eta$-$\phi$ plane. The variables are introduced in three different channels as is the case of an RGB picture, where the hadronic deposit is noted with blue, the electromagnetic deposit with green and the track transverse momentum with red. The schematic illustration of the jet image is shown in Figure \ref{fig:cnn_dsgamma}. 


\textls[-25]{Our CNN model consists of 5 layers: 3 convolutional and 2 fully connected dense layers. The number of nodes in the convolutional layers are 30, 8 and 8, respectively. The window sizes are [3 $\times$ 3] and [5 $\times$ 5] in the last layer, while the activation function is \texttt{tanh} in all three cases. A maxpooling layer is added after the second convolutional layer. The number of nodes in the first dense layers is 10 with the \texttt{ReLU} activation function. The output layer is again a dense \texttt{sigmoid}. The parameters of the final CNN model are summarised in Table \ref{tab:model}.}

\subsubsection{Combined Network}

To further improve the efficiency of the network, the DNN and the CNN models are merged into a single network. In this case, the output of the DNN and the output of the CNN are the inputs of the next combined layer. The last layer of the model performs the classification and the results depend both on the output of the CNN and the DNN.

The best performing combined network has slightly different number of nodes within the DNN layers: 33, 20 and 14, respectively. Another significant change compared to the previously introduced models is the absence of the dense layers after the convolutional layers. Instead, a combined dense layer is introduced with 8 nodes and \texttt{ReLu} activation function. The classification happens in the last \texttt{sigmoid} layer. The parameters of the combined model are summarised in the last column of Table \ref{tab:model}. 

\section{Results}\label{sec3}

\begin{figure}[H]
     \includegraphics[width=0.8\textwidth]{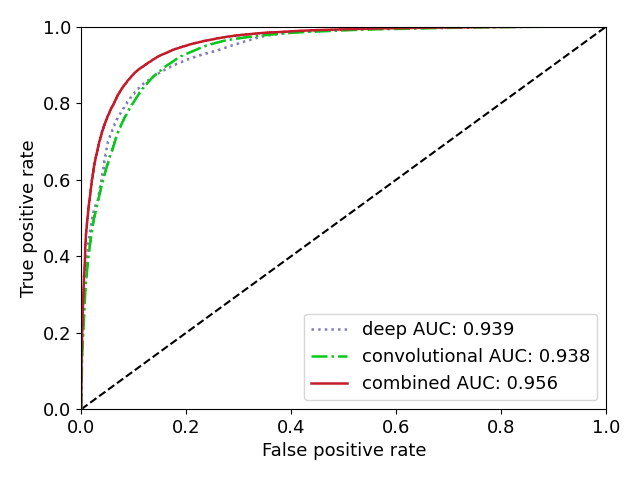}
     \caption{ROC curves for the different network types.}
     \label{fig:roc}
\end{figure}

\begin{figure}[H]
     \begin{subfigure}[b]{0.3\textwidth}
         \includegraphics[width=\textwidth]{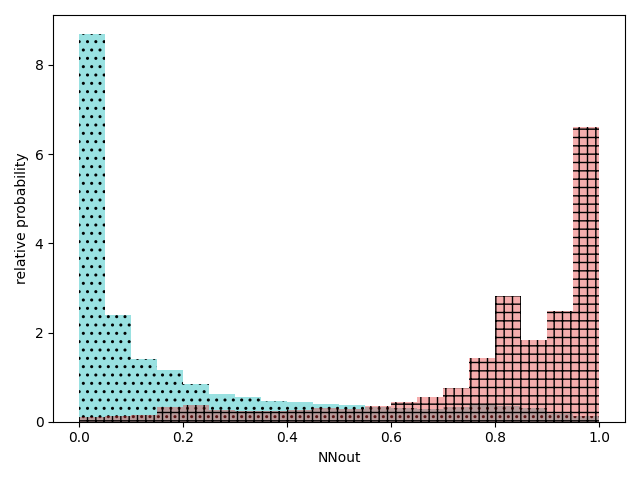}
         \caption{\centering{DNN}}
         \label{fig:nnout_dnn}
     \end{subfigure}
     \hfill
     \begin{subfigure}[b]{0.3\textwidth}
         \includegraphics[width=\textwidth]{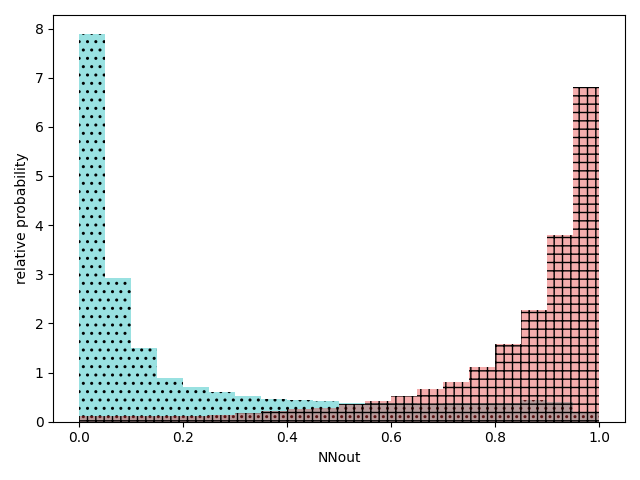}
         \caption{\centering{CNN}}
         \label{fig:nnout_ccn}
     \end{subfigure}
     \hfill
     \begin{subfigure}[b]{0.3\textwidth}
         \includegraphics[width=\textwidth]{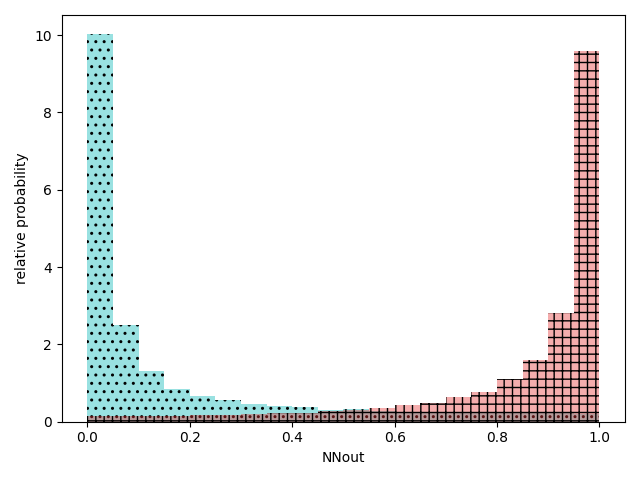}
         \caption{\centering{Combined network}}
         \label{fig:nnout_combined}
     \end{subfigure}
     \caption{Output of the different networks for background (blue dotted pattern) and signal (red square grid pattern).}
     \label{fig:nnout}
\end{figure}

The ROC curves of the different models are presented in Figure \ref{fig:roc}, while the output distributions of the models can be seen in Figure \ref{fig:nnout}. Table \ref{tab3} shows the AuC values of the different networks defined previously. As is expected, the combined model performs the best with 0.956, which corresponds to a signal efficiency of 47\% at a background rejection factor of 100 or 15\% at a background rejection factor of 1000. Using DNN only one can reach a signal efficiency of 38\% for a background rejection factor of 100 or 15\% for 1000, while using only CNN the efficiency is 35\% at 100 or 9\% at 1000 times background rejection. As it can be seen, the performance is significantly better against a single background of gluon jets than against quark jets. This can be further improved if one uses only a gluon sample for training to an AuC of 0.991.

\begin{table}[H]
    \caption{Overview 
    of the training results using the combined network. Mixed background test samples contain 50\% quark and 50\% gluon jets.}
    \centering
    \begin{tabularx}{\textwidth}{CCCC}
        \toprule
        \textbf{Network Type} & \textbf{Test Sample} & \textbf{Training Sample} & \textbf{AuC} \\
        \midrule
               DNN & \Ds vs. mixed &  \Ds vs. mixed & 0.939\\
        \midrule 
        CNN & \Ds vs. mixed &  \Ds vs. mixed & 0.938\\
        \midrule 
        \multirow{5}{*}{Combined} & \Ds vs mixed &  \Ds vs mixed & 0.956 \\
            & \Ds vs. gluon &  \Ds vs. mixed & 0.987 \\
            & \Ds vs. quark &  \Ds vs. mixed & 0.935 \\
            & \Ds vs. gluon &  \Ds vs. gluon & 0.991 \\
            & \Ds vs. quark &  \Ds vs. quark & 0.946 \\
        \bottomrule
    \end{tabularx}
    \label{tab3}%
\end{table}
\unskip
\unskip

The tagging rate of the network for various samples used and not used during the training is presented in Table~\ref{tab:diff_samples}. Here a cut-off value of 0.75 is used. We find that for charm jets the results are not materially different from the generic quark-jet sample and this indicates that the absence of fragmentation tracks around the jets and a narrow jet with low multiplicity are more important than the exact $D$-meson 
decay topology. For hadronic $\tau$ decays, we find a high tagging rate, which is not surprising, given that $\tau$ leptons are also produced in a colour-singlet state and more than 5\% of the $D_s$ mesons decay to $\tau$s.
\begin{table}[H]
    \caption{Jet 
    tagging rate for different samples. For $c\bar{c}$ and $b\bar{b}$ samples, the tagging rate is separately evaluated for events, where the jet contains a truth \Ds.}
    \begin{tabularx}{\textwidth}{LCC}
        \toprule
        \textbf{Sample}           & \multicolumn{2}{c}{\textbf{Tagging Rate}} \\
        \midrule
        \Ds$\gamma$      & \multicolumn{2}{c}{79\%}  \\
        $qq$             & \multicolumn{2}{c}{9\%}  \\
        $gg$             & \multicolumn{2}{c}{1\%}  \\
        $\tau\tau$       & \multicolumn{2}{c}{62\%}  \\
        $\Upsilon\gamma$ & \multicolumn{2}{c}{3\%}  \\
        (\JP)$\gamma$    & \multicolumn{2}{c}{16\%}  \\
        $\phi\gamma$     & \multicolumn{2}{c}{12\%}  \\
        \midrule
                         &Jet with a truth \Ds & Jet without a truth \Ds\\
        \midrule
        $c\bar{c}$        & 9\% & 7\% \\
        $b\bar{b}$        & 1\% & 3\%  \\
        \bottomrule
    \end{tabularx}
    \label{tab:diff_samples}%
\end{table}

\vspace{-6pt}

\textls[-25]{We investigate the stability of the network performance under variations of the simulation parameters. To study this, we apply the recommended variations of the \texttt{Pythia8} framework. These variations cover a range of  possible events that differ from the base simulation: variation 1 is related to the underlying event activity, variation 2 covers the jet shapes and substructure, and the three variations 3  cover the effects of initial and final state radiation. The results of the variance in the model performance is presented in Table \ref{tab4}.}

\begin{table}[H]
    \caption{Variations in the AuC for different \texttt{Pythia8} tunes.}
    \begin{tabularx}{\textwidth}{lCC}
        \toprule
        \textbf{Parameter}   &   \textbf{+Variation} & \textbf{$-$Variation}\\ 
        \midrule
        Var1: UE activity                 & $-$0.008 & 0.003 \\
        Var2: jet shapes and substructure & $-$0.001 & 0.010 \\
        Var3a: ISR/FSR $t\bar{t}$ gap     & $-$0.002 & 0.007 \\
        Var3b: ISR/FSR 3/2 jet ratio      & $-$0.011 & 0.002 \\
        Var3c: ISR                        & $-$0.007 & 0.006 \\
        \bottomrule
    \end{tabularx}

    \label{tab4}%
\end{table}

\vspace{-6pt}

The effect of pileup is also taken into account during the analysis. Within the \texttt{Delphes} framework the additional tracking and vertexing information is not available, meaning that our estimate is worse than the real life conditions on the LHC experiments. We simulated samples with a pileup of $\langle\mu\rangle = $~40 meaning on average 40 pileup interaction, which is the expected amount for LHC Run 3 conditions. The retrained network, without further optimisation, shows a drop of 0.076 in the AuC, meaning that, while pileup has a significant effect, the model is still able to identify \Ds mesons. One can note, however, that pileup mitigation techniques implemented in \texttt{Delphes} are suboptimal; hence, the expected effect with real data is smaller.

\section{Discussion}\label{sec4}

In this section prospects for the measurement of $\mathcal{B}(W\to D_{s}\gamma)$ using the method described previously are studied. For the purpose of this exercise it is assumed that low-pileup data corresponding to the integrated luminosity of 1 fb$^{-1}$ are collected during LHC Run 3. Events are required to have one jet tagged as $D_{s}$ and an isolated photon with $p_{T}>$~30~GeV. Events with invariant mass of jet-photon system  $\pm$10~GeV around $W$ boson mass are selected. Triggering efficiency is assumed to be 100\%. The optimised network cut-off of 0.75 provides the best sensitivity. Total signal efficiency for $W^{+}\to D_{s}\gamma ~(W^{-}\to D_{s}\gamma)$ is estimated to be 15.5\% (18.7\%), respectively.

\begin{figure}[H]
     \includegraphics[width=0.5\textwidth]{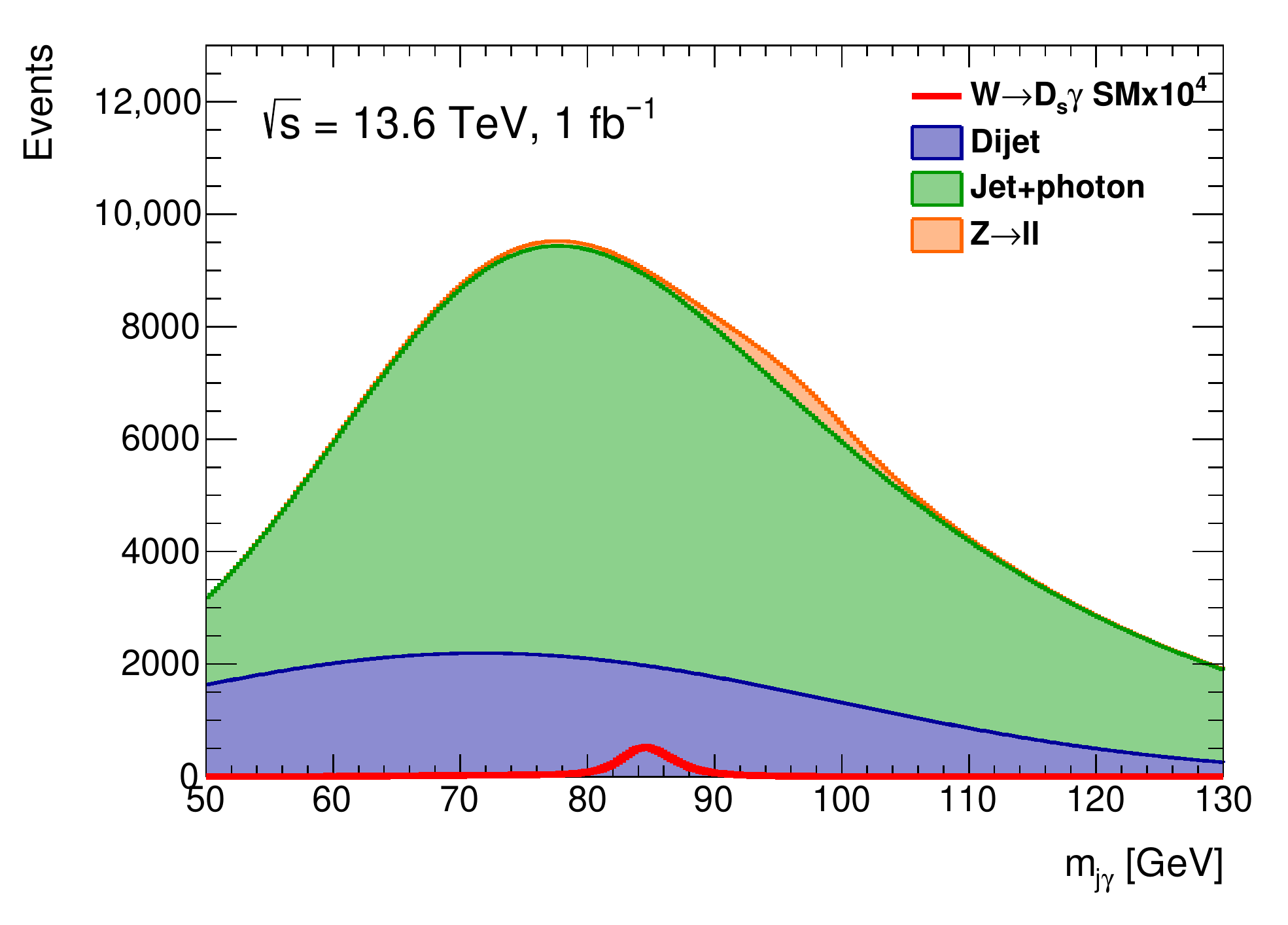}
     \caption{Distribution 
 of the 
 invariant mass of $D_{s}$ tagged jet-plus-photon system. The signal is scaled with a factor of 10$^4$.}
     \label{fig:massplot}
\end{figure}

\textls[-20]{In order to estimate background level, large MC samples of $pp\to gg$ and $pp\to qq$, as well as  $pp\to q\gamma$,  $Z\rightarrow ee$, and $Z\rightarrow \tau\tau$ are generated with \texttt{MADGRAPHv5} and \texttt{Pythia8}. The detector response is simulated via the \texttt{Delphes} package using the ATLAS detector configuration files. Backgrounds are normalised according to their generated cross-sections. The total level of background is estimated to be 930,000 events corresponding to the integrated luminosity of 1 fb$^{-1}$. The background is dominated by the QCD process while less than 1\% of the total background arises from $Z$ boson events. Figure~\ref{fig:massplot} shows the distribution of $D_{s}$ tagged jet-plus-photon invariant mass for the backgrounds and $W\to D_{s}\gamma$ signal normalised to the integrated luminosity of 1 fb$^{-1}$. The  signal histogram is overlaid and scaled by a factor of 10,000.   }

The $CL_{s}$ method~\cite{Read:2002hq,Junk:1999kv} is used to calculate the upper limit on the branching fraction of the $W\to D_{s}\gamma$ decay. The expected number of signal plus background events is
\begin{linenomath}
\begin{equation}
    N_{exp} = 
    \epsilon \sigma_{pp\to W}\mathcal{B}(W\to D_{s}\gamma) \int\mathcal{L} dt + N_{bg},
\end{equation}
\end{linenomath}
where $\epsilon$ is event selection efficiency of the signal, $\sigma_{pp\to W}$ is the inclusive production cross-section for the $W$ boson evaluated at the NNLO in QCD, $\int\mathcal{L} dt$ is the integrated luminosity, and $N_{bg}$ is the expected number of background events. Uncertainties on $\epsilon$, $\int\mathcal{L} dt$, and $N_{bg}$ are assumed to be Gaussian, and correlations between these are neglected. In this study total signal uncertainty is assumed to be 10\% and has only marginal impact on the calculated limit. The uncertainty on the background level is assumed to be 0.5\% as obtained in the ATLAS search for radiative Higgs boson decay~\cite{ATLAS:2017gko}. The upper 95\% CL (confidence level) limit on the ``signal strength'' $\sigma_{pp\to W}\mathcal{B}(W\to D_{s}\gamma)$, with production cross-section fixed, is set using Poisson statistics and the above equation.  The limit is obtained with $CL_{s} = CL_{s+b}/CL_{b} \le 0.05$, where $CL_{s+b}$ is the  confidence level for signal and background, and $Cl_{b}$ is the confidence level for the background alone.

\begin{figure}[H]
     \includegraphics[width=0.6\textwidth]{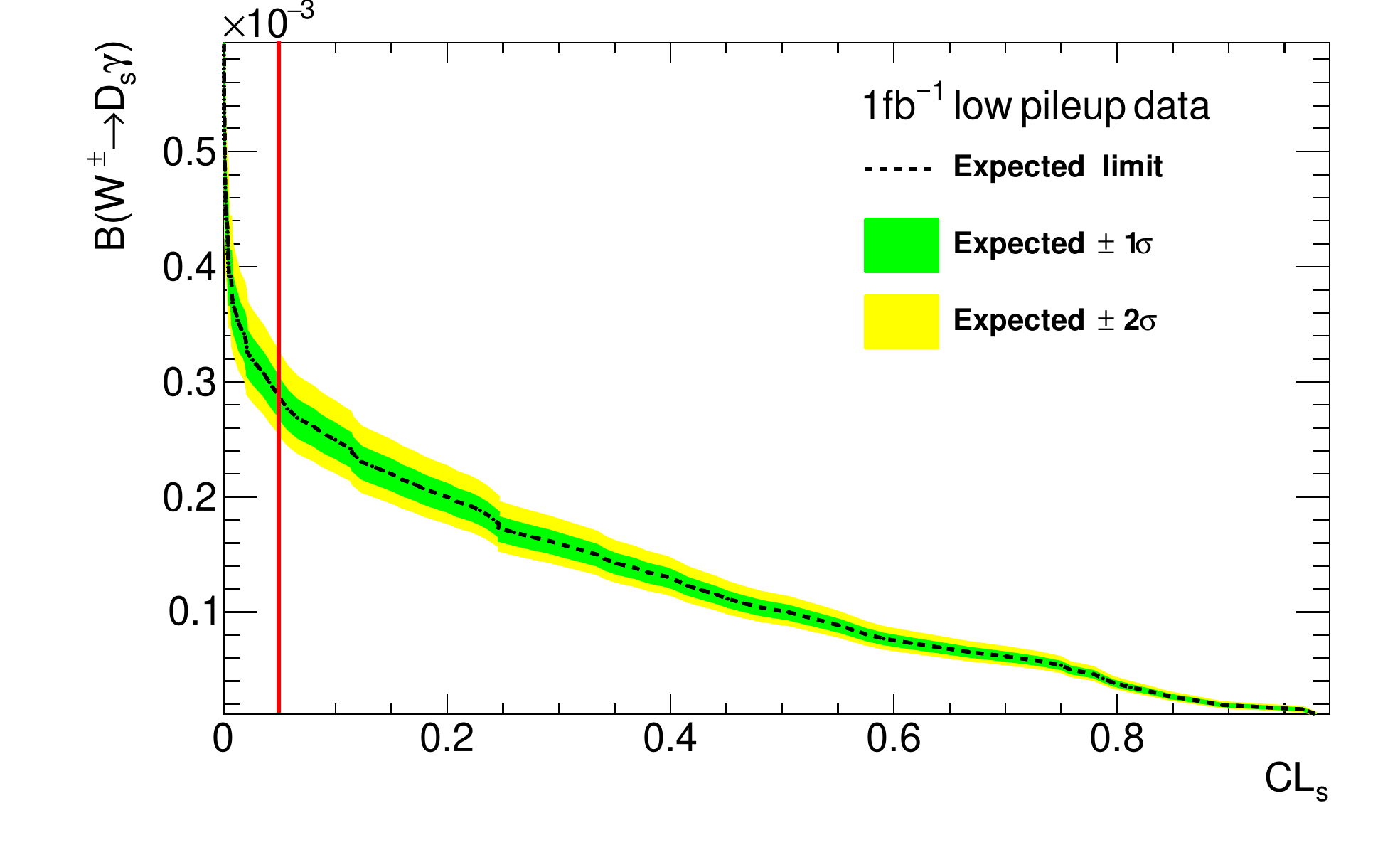}
     \caption{Expected 
 upper limit on branching fraction of the $W\to D_{s}\gamma$  decay. The vertical line corresponds to $CL_{s}~=~$0.05.  The branching fractions higher than $2.87\times10^{-4}$ are excluded at 95\% CL.
}
     \label{fig:cls}
\end{figure}

The calculated $CL_{s}$ exclusion as a function of branching fraction of  $W\to D_{s}\gamma$ is shown in Figure~\ref{fig:cls}. 

\unskip

The expected upper limit at the 95\% CL is determined to be:
\begin{linenomath}
\begin{equation}
    \mathcal{B}(W\to D_{s}\gamma) < (2.87\pm0.22)\times10^{-4},
\end{equation}
\end{linenomath}
which is by a factor of two compared to the observed upper limit from LHCb. 

With the entire Run 3 dataset corresponding to about 300 fb$^{-1}$, assuming trigger efficiency of 40\% and taking into account deterioration of the $D_{s}$ tagger due to high pileup, the expected upper limit  improves to $\mathcal{B}(W\to D_{s}\gamma) < 1.6\times10^{-4}$.
Development of a dedicated trigger is needed to achieve corresponding precision.

\section{Conclusions}

The algorithm to identify jets originating from \Ds mesons
in radiative $W$ decays presented in this paper shows a good efficiency of 47\% for signal with a 100 times rejection of jets from quarks and gluons. Against a single background of gluon jets, the algorithm works even better. The algorithm is stable under the variations of the simulation parameters and it also works in the presence of pileup but at a significant loss of performance. The algorithm opens up the possibility of further improving measurements and searches involving \Dm mesons, especially in the case of the rare decays that suffer from low statistics. We find very similar performance for a deep neural network and a convolutional
neural network, with a combined network of the two performing best. 
With a low pileup dataset corresponding to the integrated luminosity of a 1 fb$^{-1}$ upper limit on the branching fraction of  $W\to D_{s}\gamma$ decay can be determined at the level of $\mathcal{B}(W\to D_{s}\gamma) < 2.9\times10^{-4}$.

\vspace{6pt}

\authorcontributions{Conceptualization, N.d.G.; methodology, E.B., N.d.G., and N.V.; software, E.B.; validation, E.B., N.d.G., and N.V.; formal analysis, E.B.; investigation, E.B.; resources, E.B.; data curation, E.B.; writing---original draft preparation, E.B., N.d.G., and N.V.; writing---review and editing, E.B., N.d.G. and N.V.; visualization, E.B.; supervision, N.d.G. and N.V.; project administration, E.B.; funding acquisition, N.d.G., N.V.. All authors have read and agreed to the published version of the manuscript.} 


\funding{This research received no external funding.}

\dataavailability{Configuration files for data generation and analysis software can be found in github at \url{https://github.com/ebakos/DsGammaAnalysis}. Data files are available at request.}  

\acknowledgments{This work was partially supported by the Serbian Ministry of Education, Science and Technological Development and the EU Erasmus plus programme.}

\conflictsofinterest{The authors declare no conflict of interest.} 

\begin{adjustwidth}{-\extralength}{0cm}

\reftitle{References}



\PublishersNote{}
\end{adjustwidth}
\end{document}